\documentclass[a4paper,fleqn,usenatbib]{mnras}

\usepackage[T1]{fontenc}
\usepackage{ae,aecompl}

\usepackage{amsmath,amssymb,xspace} %
\usepackage{graphicx,color,paralist,times,subfig} %

\usepackage{tabularx} 
\usepackage{booktabs} 
\usepackage{units}

\newcommand{\eq}[1]{\begin{equation} #1 \end{equation}}
\newcommand{\xmm}{\textsl{XMM-Newton}\xspace} 

\title[Checking Potassium origin of new emission line at 3.5~keV] 
{Checking Potassium origin of new emission line at 3.5~keV with K~XIX line complex at 3.7~keV}

\author[D. Iakubovskyi]{
Dmytro Iakubovskyi$^{1}$\thanks{E-mail: yakubovskiy@bitp.kiev.ua}
\\
$^{1}$Bogolyubov Institute of Theoretical Physics, Metrologichna Str. 14-b, 03680, Kyiv, Ukraine\\
}

\date{Accepted XXX. Received YYY; in original form ZZZ}

\pubyear{2015}

\begin{document}
\label{firstpage}
\pagerange{\pageref{firstpage}--\pageref{lastpage}}
\maketitle

\begin{abstract}

Whether the new line at $\sim$3.5~keV, recently detected in different samples of galaxy clusters, Andromeda 
galaxy and central part of our Galaxy, is due to Potassium emission lines, is now unclear. 
By using the latest astrophysical atomic emission line database AtomDB v.~3.0.2, we show 
that the most prospective method 
to directly check its Potassium origin will be the study of
K~XIX emission line complex at $\sim$3.7~keV with future X-ray imaging spectrometers such as Soft X-ray 
spectometer
on-board \textit{Astro-H} mission or microcalorimeter on-board \textit{Micro-X} sounding rocket experiment.
To further reduce the remaining (factor $\sim 3-5$) uncertainty of the 3.7/3.5~keV ratio one should perform
more precise modeling including removal of significant spatial inhomogeneities, 
detailed treatment of background components, and further extension of the modeled energy range.
\end{abstract}



\section{Introduction}

In 2014, new emission line at $\sim$3.5~keV has been detected in X-ray spectra of various 
galaxy clusters and Andromeda galaxy~\citep{Bulbul:14a,Boyarsky:14a}. 
 Using publicly available emission line database AtomDB~v.2.0.2~\citep{Foster:12}, 
\cite{Bulbul:14a} showed that intensities of individual atomic emission lines near 3.5~keV in spectra
of different combinations of galaxy clusters are smaller by a factor $\geq 10$ 
than the detected flux of the extra line. 
\cite{Boyarsky:14a} showed that the radial profile of the new line in Perseus galaxy outskirts does not 
coincide with expected astrophysical line distribution. In addition, \cite{Boyarsky:14c} showed that the 
total expected flux from K~XVIII lines at $\sim$3.5~keV is more than an order of magnitude smaller than the 
observed new line. So \cite{Bulbul:14a,Boyarsky:14a} argued against the astrophysical line origin 
of the detected line and suggested instead the radiative decay of dark matter particles. This new hypothesis
was found consistent with subsequent detection of the 3.5~keV line in the Galactic Centre 
region~\citep{Boyarsky:14b,Lovell:14}.

An alternative viewpoint has been presented 
in~\cite{Riemer-Sorensen:14,Jeltema:14a,Jeltema:14b,Carlson:14}, 
see also~\cite{Boyarsky:14c,Bulbul:14b,Iakubovskyi:14} 
for discussion. In~\cite{Riemer-Sorensen:14}, no extra line in \xmm/EPIC spectrum of the Galactic Centre region 
is found after adding an extra line with an arbitrary normalization at 3.51~keV (the mean energy of 
K~XVIII line emission at $\sim 3.5$~keV). Based on AtomDB~v.2.0.2, 
\cite{Jeltema:14a} compared intensities of bright emission lines from both the Galactic Centre region and 
the combined galaxy cluster sample of~\cite{Bulbul:14a} and explained the observed 3.5~keV line with mild 
($\lesssim 3$) K/Ca ratio with respect to Solar value. By styding Galactic Centre and Perseus cluster 
morphologies in different 
energy bands, \cite{Carlson:14} found positive correlation of 3.45-3.60~keV band with other bands with prominent
Ar and Ca lines and no correlation with smooth dark matter profile. 
Based on these findings, \cite{Jeltema:14a,Jeltema:14b} argued that the line
at 3.5~keV in Galactic Centre, the central part of the Perseus cluster and the combined galaxy cluster dataset 
of~\cite{Bulbul:14a} is a mixture of K~XVIII lines at 3.47 and 3.51~keV and therefore may have a 
purely astrophysical origin. 

The goal of this paper is to further investigate possible Potassium origin of 
the $\sim$3.5~keV line.
If the new line is due to Potassium emission lines in multi-temperature plasma 
as suggested by~\cite{Jeltema:14a,Jeltema:14b,Carlson:14}, 
several other Potassium lines should also be excited 
at keV energies and their intensities can be predicted using the method described in Sec.~\ref{sec:methods}. 
In Sec.~\ref{sec:results} we tabulated positions and expected intensities of these lines, 
and compared them with measured line properties in Perseus cluster and Galactic Centre. 
Our conclusions are summarized in Sec.~\ref{sec:conclusions}.

\section{Methods}\label{sec:methods}

The line flux for a multi-temperature \textsc{apec}~\citep{Smith:01a} model with electron temperatures $T_{e,i}$ is 
\eq{\label{eq:F}
F [\unit{ph\ cm^{-2}\ s^{-1}}] \equiv 10^{14} \sum\limits_{i} 
\left[\epsilon(T_{e,i})\times A_i \times N_i\right],}
where $\epsilon(T_{e,i})$ is the line emissivity (in $\unit{cm^3\ s^{-1}}$), 
$A_i$ is the abundance of the line emitting element (with respect to Solar values adopted in~\cite{Anders:89}),
$N_i = \frac{10^{-14}}{4\pi D_A^2 (1+z)^2}\int n_e n_H dV$ is the \textsc{apec} model normalization 
(in cm$^{-5}$), $D_A$ is the angular diameter distance to the source 
(in cm), $z$ is the source redshift, $\int n_e n_H dV$ is the plasma emission measure (in cm$^{-3}$) of $i$-th 
thermal component. 
Using (\ref{eq:F}), one can determine the astrophysical flux of 
the new line from measured fluxes of bright astrophysical lines emitted by \emph{the other} elements. Indeed,
if one measures the flux of the bright ``reference'' line  \eq{F^{(ref)} = 10^{14} \sum\limits_{i} 
\left[\epsilon^{(ref)}(T_{e,i})\times A^{(ref)}_i N_i\right],} 
one can rewrite the flux of the new line \eq{F^{(new)} = 10^{14} \sum\limits_{i} \left[\epsilon^{(new)}
(T_{e,i})\times A^{(new)}_i N_i\right]} 
through $F^{(ref)}$
\eq{\label{eq:line-recovery-0} F^{(new)} = F^{(ref)}\times \frac{\sum\limits_{i} 
\left[\epsilon^{(new)}(T_{e,i})\times A^{(new)}_i \times N_i\right]}{\sum\limits_{i} 
\left[\epsilon^{(ref)}(T_{e,i})\times A^{(ref)}_i \times 
N_i\right]}.}

Assuming abundances of all components to be the same ($A^{(ref)}_i = A^{(ref)}$, $A^{(new)}_i = A^{(new)}$),
one obtains more convenient expression used in~\cite{Bulbul:14a,Jeltema:14a,Bulbul:14b,Jeltema:14b}
\eq{\label{eq:line-recovery} F^{(new)} = F^{(ref)}\times \frac{A^{(new)}}{A^{(ref)}}\times \frac{\sum\limits_{i} 
\left[\epsilon^{(new)}(T_{e,i})\times N_i\right]}{\sum\limits_{i} \left[\epsilon^{(ref)}(T_{e,i})\times 
N_i\right]}.} The predicted line flux $F^{(new)}$ linearly depends on 
the Potassium-to-metal abundance ratio $A^{(new)}/A^{(ref)}$ which is not measured directly 
(due to absence of strong K lines) and has to be determined further, 
e.g. from chemical evolution models~\citep{Timmes:94,Romano:10} 
or optical~\citep{Shimansky:03,Andrievsky:10} or X-ray~\citep{Phillips:15} line studies.
Possible difference of $A^{(new)}_i/A^{(ref)}_i$ ratios among components in multi-temperature plasma 
(e.g. due to supernova ejecta) not discussed in~\cite{Bulbul:14a,Jeltema:14a,Bulbul:14b,Jeltema:14b} 
will introduce further uncertainties to this method. As a result, 
it is very hard to robustly confirm or rule out the astrophysical origin of the new line. 

Following~\cite{Bulbul:14b,Jeltema:14b}, we avoid the above-mentioned uncertainty by studying emission lines of  
\emph{the same} element so that $A^{(new)}\equiv A^{(ref)}$ and (\ref{eq:line-recovery-0}) simplifies:
\eq{\label{eq:line-recovery-1} F^{(new)} = F^{(ref)}\times \frac{\sum\limits_{i} 
\left[\epsilon^{(new)}(T_{e,i})\times N_i\right]}{\sum\limits_{i} \left[\epsilon^{(ref)}(T_{e,i})\times 
N_i\right]}.}
To calculate line emissivities, we used the latest available version of astrophysical atomic emission line 
database AtomDB~v.3.0.2.
It contains much larger dataset of atomic emission lines of all elements from Hydrogen to Nickel 
(excluding Lithium and Beryllium) 
and updated information about line emissivities. Similar to AtomDB~v2.0.2 used in previous 
papers~\citep{Bulbul:14a,Jeltema:14a,Boyarsky:14c,Bulbul:14b,Jeltema:14b}, it is restricted to lines 
with emissivity $> 10^{-20} \unit{ph\ cm^3/s}$.

To account the finite energy resolution of imaging spectrometers, we broadened each emission line with 
a Gaussian shape
$$f(E) = \frac{1}{2\pi\sigma^2}\exp\left(-\frac{(E-E_0)^2}{2\sigma^2}\right),$$ where $E_0$ is 
the line position, $\sigma$ is a energy dispersion (note that, for the Gaussian line full width 
at half-maximum (FWHM) of the line is defined as $2\sigma\sqrt{2\log(2)}\approx 2.35\sigma$). 
According to Fig.~5.24 of~\cite{Iakubovskyi:13}, for the EPIC imaging spectrometers on-board \xmm\ mission 
$\sigma_{EPIC} \approx 60$~eV at $E = 3-4$~keV. For comparison, we used 
future \textit{Astro-H}/SXS spectrometer with FWHM~$\lesssim 7$~eV~\citep{Mitsuda:14}\footnote{Similar energy 
resolution will be reached in \textit{Micro-X} sounding rocket experiment~\protect\citep{Figueroa-Feliciano:15} 
planning to observe the innermost part of our Galaxy.}.
Further Doppler broadening gives $\sigma_{SXS} \lesssim 5$~eV assuming gas bulk velocities for the 
Galactic Centre and the central part of Perseus cluster 
$\lesssim 500$~km/s~\citep{Tamura:13,Kitayama:14,Koyama:14}.

\section{Results}\label{sec:results}

The main properties of Potassium emission lines above 1~keV and the adjacent emission lines of the 
other elements (S, Cl, Ar, Ca) 
are summarized in Table~\ref{tab:lines}. In addition to K~XVIII lines at 3.476, 3.496, 3.498, 
3.500 and 3.515~keV, potentially responsible for the new 3.5~keV line, it also contains two rather bright 
K~XIX lines at 3.700 and 3.706~keV surronded by rather strong Ar~XVII lines at 3.683-3.685~keV,
and two very weak ($\sim 10$ times fainter than the K~XVIII line complex at 3.51~keV) 
Potassium lines above 4~keV: K~XVIII line at 4.125~keV and K~XIX line at 4.389~keV.
Below 1~keV, all Potassium lines fall into ``line forest'' region and are strongly subdominant.

\begin{figure}
  \centering
  \includegraphics[width=0.99\linewidth]{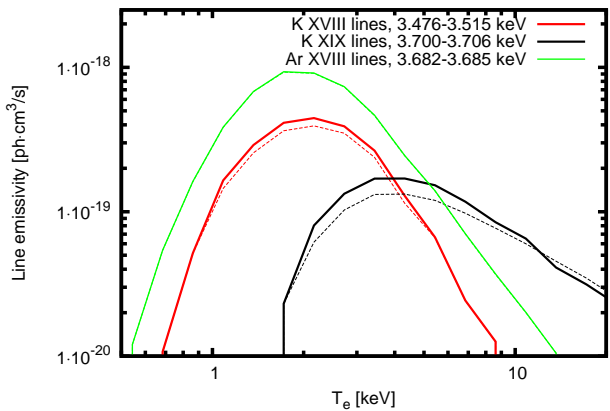}~%
  \caption{
   Emissivities of K~XVIII (3.476-3.515~keV), K~XIX (3.700-3.706~keV) and Ar~XVIII (3.683-3.685~keV) 
   line complexes as function of electron temperature $T_e$. 
   The K and Ar abundances are set to Solar~\protect\citep{Anders:89}. 
   Solid and dashed lines show contributions from AtomDB~v3.0.2
   and AtomDB~v2.0.2, respectively (no difference between these AtomDB versions for Ar~XVIII, 
  $< 15\%$ difference for K~XVIII and $< 30\%$ difference for K~XIX).
  }
  \label{fig:K-lines-atomdb302}
\end{figure}

The emissivities of K~XVIII (3.476-3.515~keV), K~XIX (3.700-3.706~keV) and Ar~XVII (3.683-3.685~keV)
line complexes as fuctions of electron temperature $T_e$ are shown in Fig.~\ref{fig:K-lines-atomdb302},
see also Fig.~8 of~\cite{Urban:14} for AtomDB v.~2.0.2.
For $T_e \gtrsim 3$~keV, 3.700-3.706~keV K~XIX line signal becomes comparable or stonger 
than that of 3.476-3.515~keV K~XVIII lines which should lead to significant detection of $\sim$3.7 keV line.

Fig.~\ref{fig:K-lines-atomdb302} shows that the 3.7/3.5~keV line ratio strongly depends on 
exact temperatures of individual components in thermal plasma. We illustrate this uncertainty with different
models parameters of the Galactic Centre and Perseus cluster spectra. 
\cite{Jeltema:14a} considered three models of Galactic Centre region: 
two-component with $T_{e,1} = 1$~keV, $T_{e,2} = 7$~keV and $N_1/N_2 = 4$; 
two-component  with $T_{e,1} = 0.8$~keV, $T_{e,2} = 8$~keV and $N_1/N_2 = 3$;
three-component with $T_{e,1} = 0.8$~keV, $T_{e,1} = 2$~keV, $T_{e,1} = 8$~keV and 
$N_1 : N_2 : N_3 = 0.17 : 1.0 : 0.075$. The corresponding 3.7/3.5~keV line ratios are 0.084,
0.189 and 0.276, respectively. \cite{Bulbul:14a} modeled MOS spectra of Perseus cluster and found $T_{e,1} = 
3.6\pm 0.6$~keV, $N_1 = 15.7\pm 7.8 \times 10^{-2}~\unit{cm}^{-5}$, $T_{e,2} = 7.6\pm 0.7$~keV, 
$N_2 = 44.0\pm 6.8 \times 10^{-2}~\unit{cm}^{-5}$ which gives 3.7/3.5~keV line ratio 
$1.77^{+4.9}_{-0.46}$.

\begin{figure*}
  \centering
  \includegraphics[width=0.45\textwidth]{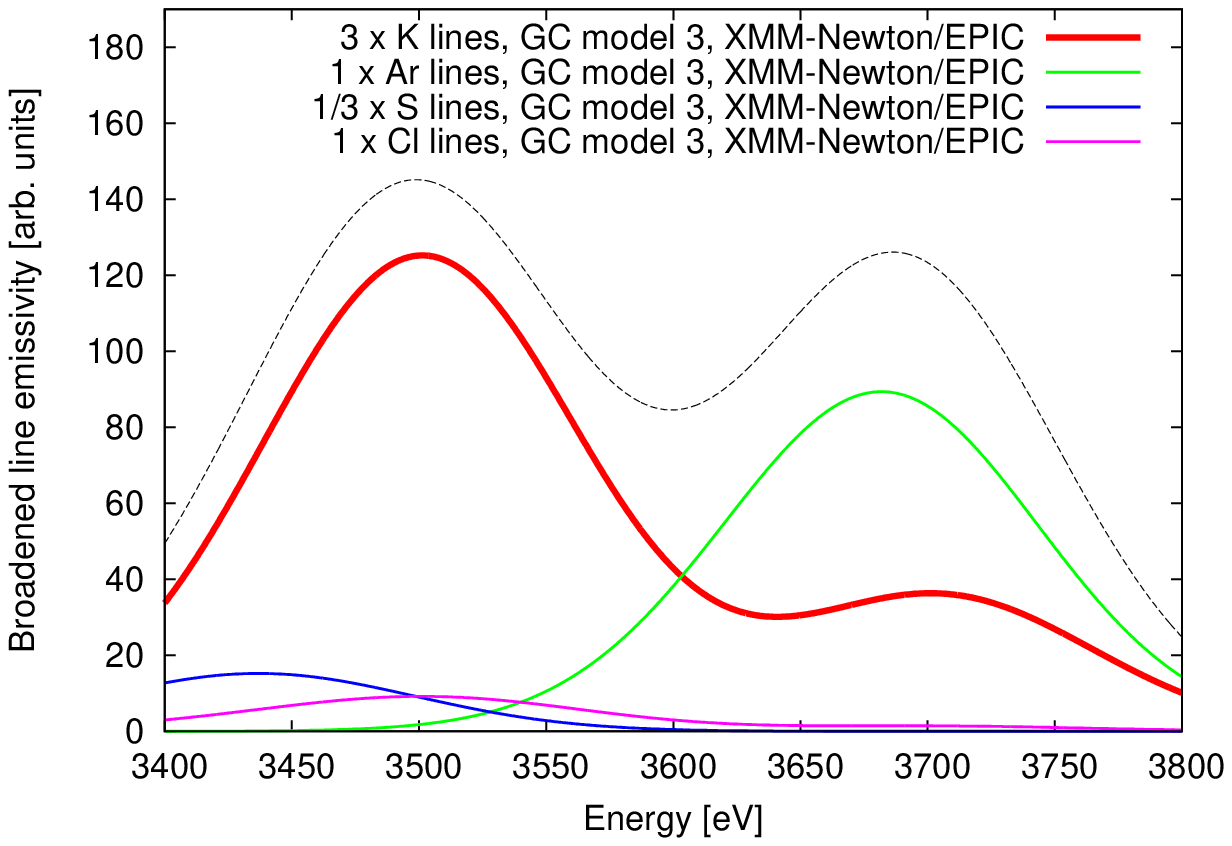}~%
  \includegraphics[width=0.45\textwidth]{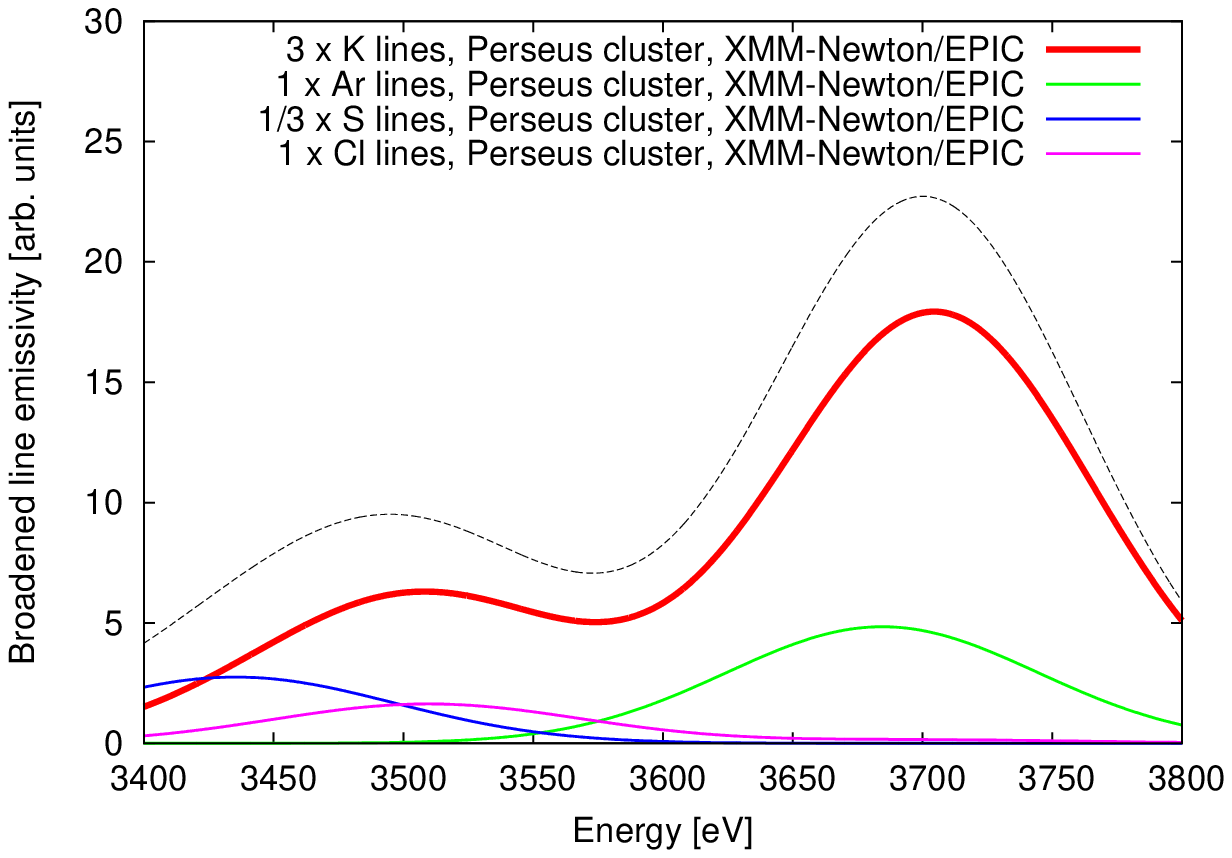}
  \caption{\textit{Left:} Broadened with $\sigma_{XMM} = 60$~eV line emissivities as functions of energy 
  for three-component model of~\protect\cite{Jeltema:14a} of Galactic Centre. The relative S, Ar, Cl and K 
  abundances are set to 1/3 : 1 : 1 : 3, see Sec.~2.2 of~\protect\cite{Jeltema:14a}. 
  Thin dashed line shows the total emissivity.
  \textit{Right}: Broadened line emissivities as functions of energy for the best-fit 
  two-component model of Perseus cluster given by~\protect\cite{Bulbul:14a}.
  The relative element abundances in Solar units given by~\protect\cite{Anders:89} are the same as in the left Figure.}
  \label{fig:K-lines-3400-3800-xmm}
\end{figure*}

\begin{figure*}
  \centering
  \includegraphics[width=0.45\textwidth]{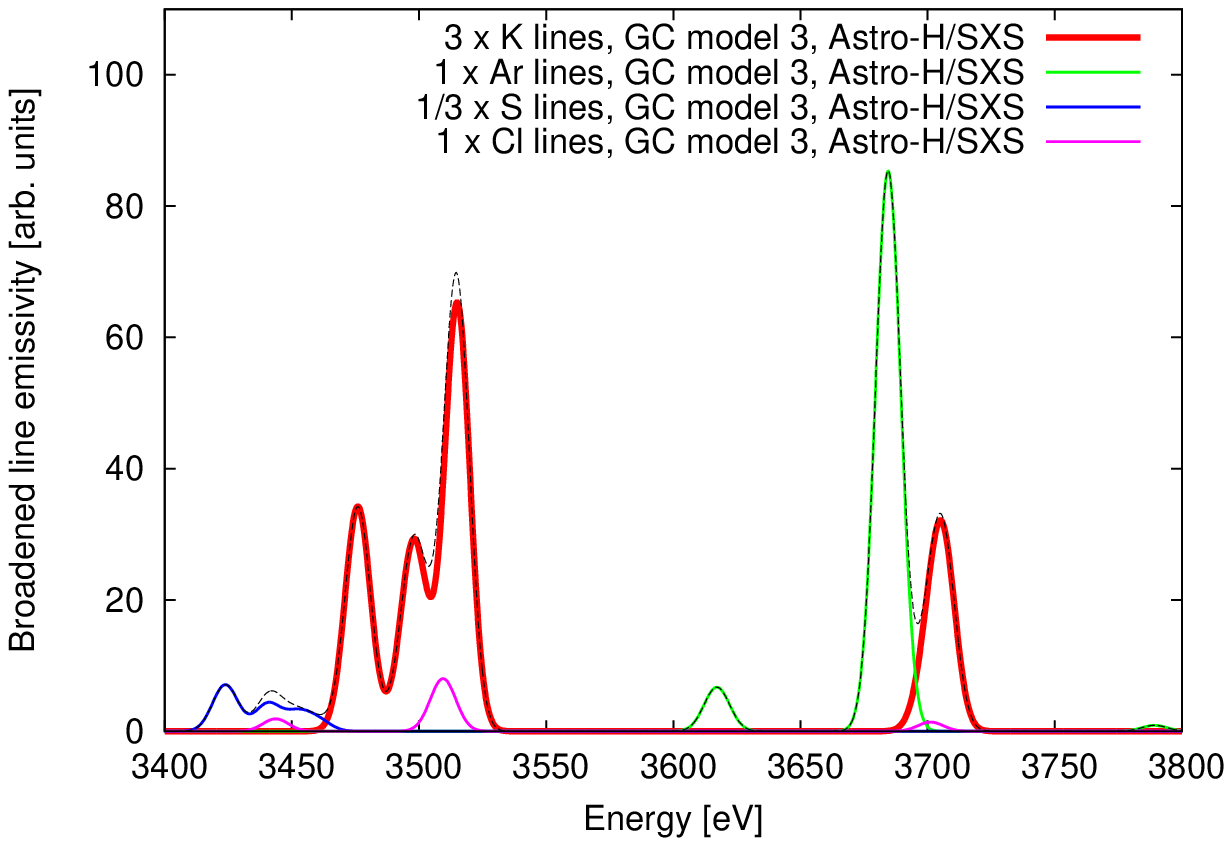}~%
  \includegraphics[width=0.45\textwidth]{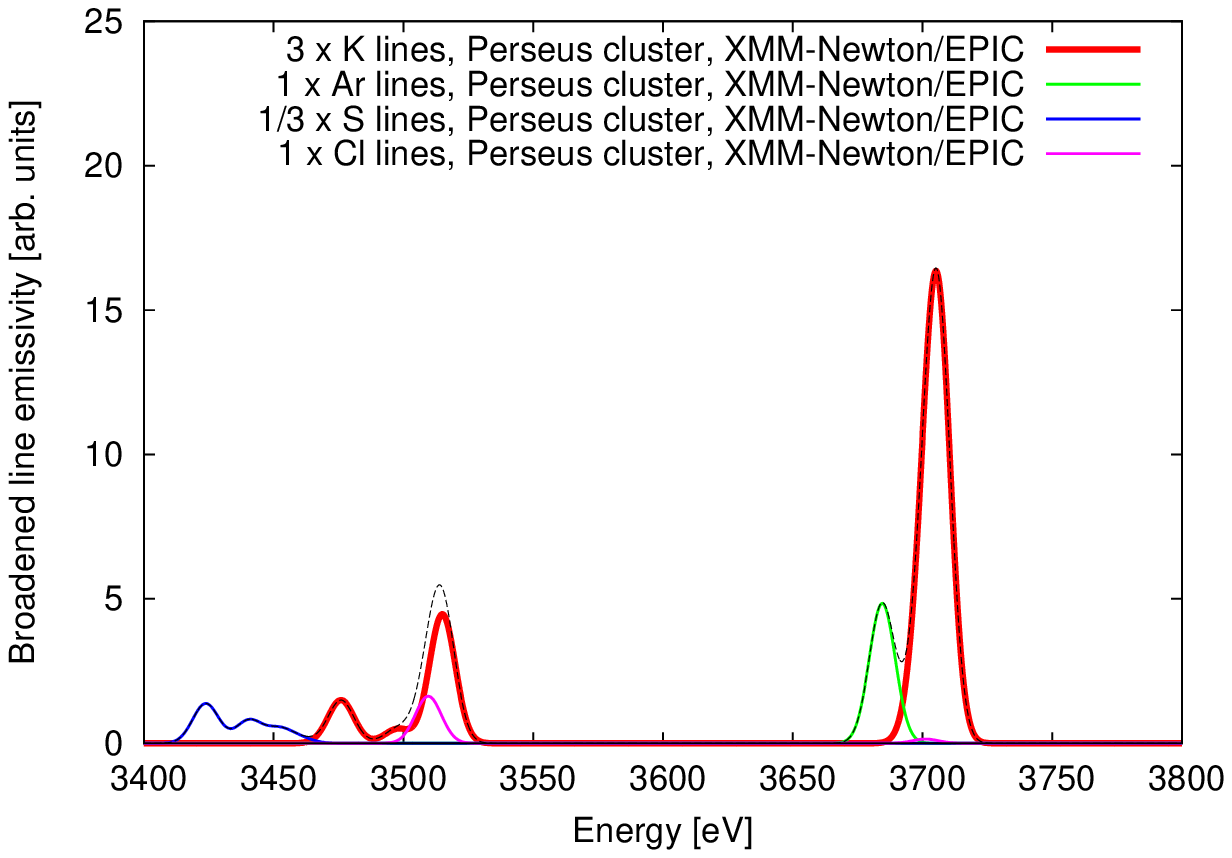}
  \caption{The same as in the previous Figure but broadened with \textit{Astro-H}/SXS or \textit{Micro-X} 
  energy resolution $\sigma_{SXS} = 5$~eV (see text).}
  \label{fig:K-lines-3400-3800-sxs}
\end{figure*}

In addition to the factor $\sim$3-5 uncertainty in 3.7/3.5~keV line ratio for the Galactic Center and 
Perseus cluster, another uncertainty comes from strong contribution of Ar~XVII lines at 3.683-3.685~keV 
located only $\sim$20~eV away from 3.700-3.706~keV K~XIX line complex. 
We visualize the relative contributions of these lines by
broadening line emissivities for Galactic Centre (three-component model of~\cite{Jeltema:14a}) 
and Perseus cluster (best-fit two-component model of~\cite{Bulbul:14a} for 
combined MOS spectrum) with \xmm/EPIC energy resolution, see Fig.~\ref{fig:K-lines-3400-3800-xmm}
for details. In both sources, the S, Ar, Cl and K 
abundances ratio are set to 1/3 : 1 : 1 : 3, close to that obtained by~\cite{Jeltema:14a} for 
Galactic Centre region. This uncertainty can be avoided by using future instruments with higher energy resolution,
see Fig.~\ref{fig:K-lines-3400-3800-sxs} for the planned \textit{Astro-H}/SXS instrument and \textit{Micro-X} 
sounding rocket experiment.

Existing stacked X-ray spectra of nearby galaxies and galaxy groups~\citep{Iakubovskyi:13,Anderson:14} 
did not allow to draw any conclusion about the astrophysical origin of the 3.5~keV line. Indeed,
a simple (but abundance-dependent) estimate of Ar~XVIII lines strength at 3.683-3.685~keV (dominating at 
low electron temperatures $T_e < 1$~keV expected from~\cite{Anderson:14} according to 
Fig.~\ref{fig:K-lines-atomdb302}) can be made by using luminous S~XV line complex at $\sim$2.45~keV, 
see e.g. Fig.~3 of \cite{Bulbul:14b} for AtomDB 2.0.2. 
An estimate with the newest AtomDB 3.0.2 gives the expected 2.45keV/3.68keV line flux ratio 
$> 170 \times \text{Abund[S]}/\text{Abund[Ar]}$ for $T_e < 1$~keV. 
More precise (abundance-independent) estimate can be made using the brightest Ar~XVII line complex at 
3.10-3.14 keV which gives 3.14keV/3.68keV line flux ratio $> 20$ for $T_e < 1$~keV.

\begin{table*}
\centering
\begin{tabular}[c]{lccccc}
\hline
 Position [eV]~~~ & Ion & Transition & T$_e$ range [keV] & T$_{e,max}$ [keV] & $\epsilon_{max}/\epsilon_{2.16}/\epsilon_{4.32}$ [$10^{-20} \unit{ph\ cm^3/s}$] \\  
\hline
3423.78 & S~XVI & $38 \to 1$ & 1.08--10.8 & 2.16 & 6.8/6.8/3.9 \\
3424.84 & S~XVI & $39 \to 1$ & 1.08--21.6 & 2.16 & 13.8/13.8/9.4 \\
3440.48 & S~XVI & $51 \to 1$ & 1.37--8.62 & 2.16 & 3.9/3.9/2.2 \\
3440.52 & S~XVI & $52 \to 1$ & 1.08--13.7 & 2.16 & 8.7/8.7/6.0 \\
3443.61 & Cl~XVI & $31 \to 1$ & 1.08--2.72 & 1.72 & 3.0/2.4/--- \\
3451.92 & S~XVI & $66 \to 1$ & 1.37--6.84 & 2.16 & 2.9/2.9/1.9 \\
3451.95 & S~XVI & $67 \to 1$ & 1.37--10.8 & 2.16 & 5.1/5.1/3.4 \\
3460.11 & S~XVI & $83 \to 1$ & 1.72--5.44 & 2.16 & 2.1/2.1/1.4 \\
3460.13 & S~XVI & $84 \to 1$ & 1.37--6.84 & 2.16 & 3.0/3.0/2.0 \\
\textbf{3475.94} & \textbf{K~XVIII} & $2 \to 1$ & \textbf{0.862--5.44} & \textbf{2.16} & \textbf{11.8/11.8/3.5} \\ 
\textbf{3496.43} & \textbf{K~XVIII} & $4 \to 1$ & \textbf{1.08--3.43} & \textbf{2.16} & \textbf{3.4/3.4/---} \\ 
\textbf{3497.78} & \textbf{K~XVIII} & $5 \to 1$ & \textbf{1.08--4.32} & \textbf{2.16} & \textbf{4.8/4.8/1.3} \\ 
\textbf{3500.24} & \textbf{K~XVIII} & $6 \to 1$ & \textbf{1.08--3.43} & \textbf{2.16} & \textbf{2.7/2.7/---} \\ 
3508.66 & Cl~XVII & $5 \to 1$ & 1.72--6.84 & 2.72 & 2.6/2.3/2.0 \\
3509.79 & Cl~XVII & $8 \to 1$ & 1.37--13.7 & 2.72 & 5.8/5.0/3.9 \\
\textbf{3514.99} & \textbf{K~XVIII} & $7 \to 1$ & \textbf{0.684--8.62} & \textbf{2.16} & \textbf{21.9/21.9/8.1} \\ 
3521.93 & Cl~XVI & $41 \to 1$ & 1.72--1.72 & 1.72 & 1.1/---/--- \\
3616.91 & Ar~XVI & $10078 \to 3$ & 0.544--3.43 & 1.37 & 9.4/5.0/--- \\
3617.65 & Ar~XVI & $10077 \to 2$ & 0.684--2.72 & 1.37 & 3.6/1.9/--- \\
3682.78 & Ar~XVII & $11 \to 1$ & 1.08--3.43 & 1.72 & 3.8/3.4/--- \\
3684.52 & Ar~XVII & $13 \to 1$ & 0.544--13.7 & 1.72 & 89.5/88.0/24.5 \\
\textbf{3700.08} & \textbf{K~XIX} & $3 \to 1$ & \textbf{2.16--10.8} & \textbf{3.43} & \textbf{3.8/1.9/3.7} \\ 
3701.19 & Cl~XVII & $13 \to 1$ & 2.16--4.32 & 2.72 & 1.5/1.3/1.1 \\
\textbf{3705.97} & \textbf{K~XIX} & $4 \to 1$ & \textbf{1.72--34.3} & \textbf{4.32} & \textbf{13.3/6.2/13.3} \\ 
3787.74 & Ar~XVI & $10125 \to 3$ & 1.37--1.37 & 1.37 & 1.0/---/--- \\
3788.9 & Ar~XVI & $10126 \to 2$ & 1.08--1.72 & 1.08 & 1.2/---/--- \\
3789.25 & Ar~XVI & $10127 \to 3$ & 0.862--2.16 & 1.08 & 2.3/1.3/--- \\
 \hline
4100.12 & Ca~XX & $3 \to 1$ & 1.37--68.4 & 4.32 & 109.7/28.9/109.7 \\
4105.44 & Ca~XIX & $10067 \to 23$ & 2.16--5.44 & 3.43 & 1.9/1.2/1.6 \\
4105.57 & Ca~XIX & $10027 \to 13$ & 2.16--5.44 & 3.43 & 2.4/1.5/2.0 \\
4105.84 & Ca~XIX & $10021 \to 9$ & 2.72--4.32 & 3.43 & 1.2/---/1.0 \\
4106.25 & Ca~XIX & $10118 \to 37$ & 3.43--3.43 & 3.43 & 1.1/---/--- \\
4107.48 & Ca~XX & $4 \to 1$ & 1.37--68.4 & 4.32 & 215.2/56.5/215.2 \\
4109.93 & Ca~XIX & $10092 \to 30$ & 2.16--4.32 & 3.43 & 1.7/1.0/1.4 \\
4110.88 & Ca~XIX & $10037 \to 17$ & 2.72--4.32 & 3.43 & 1.3/---/1.0 \\
\textbf{4125.07} & \textbf{K~XVIII} & $17 \to 1$ & \textbf{1.37--4.32} & \textbf{2.16} & \textbf{2.7/2.7/1.1} \\ 
4149.72 & Ar~XVIII & $11 \to 1$ & 1.72--21.6 & 3.43 & 8.6/5.4/7.9 \\
4150.33 & Ar~XVIII & $12 \to 1$ & 1.37--34.3 & 3.43 & 18.5/11.6/16.9 \\
4249.36 & Ar~XVIII & $18 \to 1$ & 1.72--10.8 & 3.43 & 3.8/2.4/3.0 \\
4249.67 & Ar~XVIII & $19 \to 1$ & 1.72--21.6 & 3.43 & 8.3/5.2/7.7 \\
4304.59 & Ar~XVIII & $27 \to 1$ & 2.16--6.84 & 3.43 & 2.0/1.4/1.8 \\
4304.77 & Ar~XVIII & $28 \to 1$ & 1.72--13.7 & 3.43 & 4.5/2.8/4.2 \\
4337.19 & Ar~XVIII & $39 \to 1$ & 2.72--4.32 & 3.43 & 1.2/---/1.1 \\
4337.3 & Ar~XVIII & $39 \to 1$ & 2.16--8.62 & 2.72 & 2.4/1.7/2.1 \\
4358.41 & Ar~XVIII & $52 \to 1$ & 2.72--5.44 & 3.43 & 1.5/---/1.4 \\
4372.88 & Ar~XVIII & $67 \to 1$ & 3.43--3.43 & 3.43 & 1.0/---/--- \\
\textbf{4388.76} & \textbf{K~XIX} & $8 \to 1$ & \textbf{2.72--8.62} & \textbf{4.32} & \textbf{1.8/---/1.8} \\ 
\hline
\end{tabular}
\caption{Emission lines at 3.4-3.8 and 4.1-4.5~keV range from AtomDB~v3.0.2 database. 
 Level transitions are
 in AtomDB format. 
 Range of T$_e$ shows when the emissivity of a given line is $> 10^{-20}
\unit{ph\ cm^3/s}$. T$_{e,max}$ and $\epsilon_{max}$ indicate the temperature at which maximal emissivity of 
the line is reached, and the maximal value of the line emissivity; $\epsilon_{2.16}$ and $\epsilon_{4.32}$
are the line emissivities at T$_e$ = 2.16 and 4.32~keV.}
\label{tab:lines}
\end{table*}

\section{Conclusions}\label{sec:conclusions}

Determination of Potassium origin of the new line at $\sim$3.5~keV in thermal multi-component plasma 
suffers from several uncertainties, the most significant of them are the uncertainties of the plasma 
temperatures and of the relative element abundances among the thermal components.
Future study of K~XIX emission lines at $\sim$3.7~keV with high-resolution imaging spectometers such as
SXS spectrometer on-board \textit{Astro-H} mission~\citep{Mitsuda:14} or microcalorimeter on-board 
\textit{Micro-X} sounding rocket experiment~\citep{Figueroa-Feliciano:15} can completely avoid the second 
uncertainty\footnote{If the new line position resolved by the future spectrometers falls 
\emph{off} K~XVIII lines that will be the clear signature of its \emph{non-Potassium} origin.}. 

To reduce the remaining uncertainty, one has to improve the thermal 
continuum modeling including removal of significant spatial inhomogeneities, 
detailed treatment of background components (such as instrumental background and residual soft proton flares
for \xmm/EPIC), and further extension of the modeled energy range 
(possibly including the data from other instruments).

\section*{Acknowledgements}

The author thanks Michael E. Anderson for his comment about the detection of astrophysical lines in 
stacked galaxy spectra.
This work was supported by part by the SCOPE grant IZ7370-152581 and 
the Program of Cosmic Research of the National Academy of 
Sciences of Ukraine.

\let\jnlstyle=\rm\def\jref#1{{\jnlstyle#1}}\def\aj{\jref{AJ}}
  \def\araa{\jref{ARA\&A}} \def\apj{\jref{ApJ}\ } \def\apjl{\jref{ApJ}\ }
  \def\apjs{\jref{ApJS}} \def\ao{\jref{Appl.~Opt.}} \def\apss{\jref{Ap\&SS}}
  \def\aap{\jref{A\&A}} \def\aapr{\jref{A\&A~Rev.}} \def\aaps{\jref{A\&AS}}
  \def\azh{\jref{AZh}} \def\baas{\jref{BAAS}} \def\jrasc{\jref{JRASC}}
  \def\memras{\jref{MmRAS}} \def\mnras{\jref{MNRAS}\ }
  \def\pra{\jref{Phys.~Rev.~A}\ } \def\prb{\jref{Phys.~Rev.~B}\ }
  \def\prc{\jref{Phys.~Rev.~C}\ } \def\prd{\jref{Phys.~Rev.~D}\ }
  \def\pre{\jref{Phys.~Rev.~E}} \def\prl{\jref{Phys.~Rev.~Lett.}}
  \def\pasp{\jref{PASP}} \def\pasj{\jref{PASJ}} \def\qjras{\jref{QJRAS}}
  \def\skytel{\jref{S\&T}} \def\solphys{\jref{Sol.~Phys.}}
  \def\sovast{\jref{Soviet~Ast.}} \def\ssr{\jref{Space~Sci.~Rev.}}
  \def\zap{\jref{ZAp}} \def\nat{\jref{Nature}\ } \def\iaucirc{\jref{IAU~Circ.}}
  \def\aplett{\jref{Astrophys.~Lett.}}
  \def\apspr{\jref{Astrophys.~Space~Phys.~Res.}}
  \def\bain{\jref{Bull.~Astron.~Inst.~Netherlands}}
  \def\fcp{\jref{Fund.~Cosmic~Phys.}} \def\gca{\jref{Geochim.~Cosmochim.~Acta}}
  \def\grl{\jref{Geophys.~Res.~Lett.}} \def\jcp{\jref{J.~Chem.~Phys.}}
  \def\jgr{\jref{J.~Geophys.~Res.}}
  \def\jqsrt{\jref{J.~Quant.~Spec.~Radiat.~Transf.}}
  \def\memsai{\jref{Mem.~Soc.~Astron.~Italiana}}
  \def\nphysa{\jref{Nucl.~Phys.~A}} \def\physrep{\jref{Phys.~Rep.}}
  \def\physscr{\jref{Phys.~Scr}} \def\planss{\jref{Planet.~Space~Sci.}}
  \def\procspie{\jref{Proc.~SPIE}} \let\astap=\aap \let\apjlett=\apjl
  \let\apjsupp=\apjs \let\applopt=\ao \def\jcap{\jref{JCAP}}

\bsp	
\label{lastpage}
\end{document}